\documentclass[structabstract]{aa}
\usepackage{graphicx}
\usepackage{natbib}
\usepackage{amsmath,amssymb,graphicx}
\begin{document}
   \title{%A local mid-infrared investigation of the
   The scaling relation between the mass of supermassive
    black holes and the kinetic energy of random motions of the host galaxies}

%   \subtitle{}

\titlerunning{The $M_{\bullet}-M_{\mathrm{G}}\sigma^2$ relation for local galaxies}
   \author{
          L. Mancini \inst{1,2} %\thanks{}
          \and
          A. Feoli\inst{3}}
       \institute{Max-Planck-Institute for Astronomy, K\"{o}nigstuhl 17, 69117, Heidelberg, Germany \\
             \email{mancini@mpia.de}
         \and
                Department of Physics, University of Salerno, Via Ponte Don Melillo, 84084--Fisciano (SA),
                Italy\\
         \and
    Department of Engineering, University of Sannio, Piazza Roma 21, 82100--Benevento, Italy\\
              \email{feoli@unisannio.it}}
   \date{Received ; Accepted }
% \abstract{}{}{}{}{}
% 5 {} token are mandatory
  \abstract
{Thanks to the improved angular resolution of modern telescopes
and kinematic models, the existence of supermassive black holes
(SMBHs) in the inner part of galaxies, regardless their morphology
and nuclear activity, has been established on quite solid grounds.
A possible correlation between the mass of SMBHs ($M_{\bullet}$)
and the evolutionary state of their host galaxies is expected and
is currently under a heated debate.}
{Based on the recent 2D decomposition of 3.6 $\mu$m
\textit{Spiter}/IRAC images of local late- and early-type galaxies
with $M_{\bullet}$ measurements, we investigated various scaling
laws, studying what the best predictor of the mass of the central
black holes is. In particular, we focused on the
$M_{\bullet}-M_{\mathrm{G}}\sigma^2$ law, that is the relation
between the mass of SMBHs and the kinetic energy of random motions
of the corresponding host galaxies, $M_{\mathrm{G}}$ is the mass
and $\sigma$ the velocity dispersion of the host galaxy (bulge).}
{In order to find the best fit for each of the scaling laws
examined, we performed a least--squares regression of
$M_{\bullet}$ on $x$ for the considered sample of galaxies, $x$
being a whatever known parameter of the galaxy bulge. For this
purpose, we made use of both the linear regression LINMIX\_ERR and
FITEXY methods.}
{Our analysis shows that $M_{\bullet}-M_{\mathrm{G}}\sigma^2$ law
fits the examined experimental data successfully as much as the
other known scaling laws (all correlations have similar intrinsic
scatters within the errors) and shows a value of $\chi^2$
(estimated by FITEXY) better than the others, a result which is
consistent with previous determinations at shorter wavelengths.
This means that a combination of $\sigma$ and $M_{\mathrm{G}}$ (or
$R_{\mathrm{e}}$) could be necessary to drive the correlations
between $M_{\bullet}$ and other bulge properties. This issue has
been investigated by a careful, although not fully conclusive,
analysis of the residuals of the various relations.}
{In order to avoid rushed conclusions on galaxy activity and
evolution, the indirect inferring of the masses of the
supermassive black holes from the kinetic energy of random motions
via the $M_{\bullet}-M_{\mathrm{G}}\sigma^2$ relation should be
considered, especially when applied to higher redshift galaxies
($z>0.01$). This statement is suggested by a reanalysis of the
Sloan Digital Sky Survey (SDSS) data used to study the black hole
growth in the nearby Universe. By adopting the
$M_{\bullet}-M_{\mathrm{G}}\sigma^2$ relation instead of the
$M_{\bullet}-\sigma$ relation, a radio--quiet/radio--loud
dichotomy appears in the SMBH mass distribution of the
corresponding SDSS early--type AGN galaxies.}


\maketitle

%
%________________________________________________________________

%%%%%%%%%%%%%%%%%%%%%%%%%%%%%%%%%%%%%%%%%%%%%%%%%%%%%%
\section{Introduction}
%%%%%%%%%%%%%%%%%%%%%%%%%%%%%%%%%%%%%%%%%%%%%%%%%%%%%%
Scaling laws in galaxies are important to describe the mechanisms
for the initial formation of the first galaxies as well as for
their cosmic evolution. They also apply quite well to high
redshifts, implying that the interaction and merging processes
must, at least on average, preserve them \citep{schneider06}.

One of the recent successes in extragalactic astronomy was the
discovery that both early-- and late--type galaxies, close ($<100$
Mpc) to the Milky Way,  host a supermassive black hole (SMBH;
$M_{\bullet}>10^6 M_{\sun}$) at their center
\citep{kormendy95,richstone98}. Subsequently, based on more and
more precise experimental data, it was possible to draw up many
galaxy data sets, containing measures of the SMBH masses as well
as the structural parameters of the host galaxies (bulges)
\citep{magorrian98,tremaine02,marconi03,gebhardt03,haring04,aller07,graham08a,graham08b,hu08,kisaka08,feoli09,gultekin09a}.

Thanks to these catalogues, the astrophysical community identified
a large number of scaling laws, in which the mass of SMBHs
correlates with several properties of the host spheroidal
component\footnote{Here we use the terms \emph{bulge} or
\emph{spheroid} to mean the spheroidal component of a
spiral/lenticular galaxy or a full elliptical galaxy.}, such as
for instance bulge luminosity, mass, effective radius, central
potential, dynamical mass, concentration, Sersic index, binding
energy, etc.
\citep{richstone95,magorrian98,ferrarese00,gebhardt00,laor01,merritt01,%
wandel02,tremaine02,marconi03,haring04,graham05,feoli05,aller07,hopkins07b}.
Another scaling law has been recently proposed, the $M_{\bullet} -
R_{\mathrm{e}} \sigma^3$ law \citep{feoli11b}, which is based on a
pure theoretical framework, and is on the wake of the numerical
results of \citet{hopkins07a}.

Relations between $M_{\bullet}$ and X-ray luminosity,
radioluminosity \citep{gultekin09b}, momentum parameter
\citep{soker10}, number of globular clusters
\citep{burkert10,snyder11} have also been presented, whereas the
correlation with the dark matter halo is still a matter of
controversy
\citep{ferrarese02,baes03,kormendy11,graham11,volonteri11,bellovary11}.

All these scaling laws have led to the belief that SMBH growth and
bulge formation regulate each other \citep{ho04}, even if it is
difficult to understand the fundamental nature of the correlations
between SMBHs and host properties \citep{jahnke11}, also because
all such relations depend critically on the accuracy of the
published error estimates in all quantities under consideration
\citep{novak06,lauer07}.

Just as the Faber--Jackson relation, the ``traditional'' relation
between the SMBH mass and bulge velocity dispersion $\sigma$ or
stellar mass $M_{\star}$ should be projections of the same
fundamental plane relation. Current observations require a
correlation of the form $M_{\bullet} \propto
M_{\star}^{\alpha}\sigma^{\beta}$ over a simple correlation with
either $\sigma$ or $M_{\star}$ at $\geq 3 \sigma$ confidence
\citep{hopkins07a,hopkins08,marulli08}.

Actually, another competitive correlation, as opposed to the
popular $M_{\bullet}-M_{\star}$ and $M_{\bullet}-\sigma$, between
$M_{\bullet}$ and the kinetic energy of random motions of the
corresponding bulges, i.e. $M_{\star} \sigma^{2}$, has been
advanced \citep{feoli05,feoli09}. This relation has also a plausible
physical interpretation that resembles the H--R diagram: the mass
of the central SMBH, just like entropy, can only increase with
time or at most remain the same but never decrease; $M_{\bullet}$
is therefore related to the age of the galaxy. On the other hand,
the kinetic energy of the stellar bulges directly determines the
temperature of the galactic system. The goodness of the
$M_{\bullet}-M_{\star} \sigma^{2}$ relation, as a predictor of the
SMBH mass in the center of galaxies, has been already tested, with
clear positive results, over three independent galaxy samples and
in the framework of the $\Lambda$CDM cosmology, using two galaxy
formation models based on the Millennium Simulation, one by
\citet{bower06} (the Durham model) and the other by
\citet{delucia07} (the MPA model) \citep{feoli11a}.

Recently, \citet{sani11} have presented a mid-infrared investigation of
the scaling relations between SMBH masses and some of structural
parameters (luminosity, mass, effective radius, velocity
dispersion) of the host spheroids in local galaxies, based on
\textit{Spitzer}/IRAC 3.6 $\mu$m images of 57 galaxies of
different morphological types with $M_{\bullet}$ measurements.
Their results were consistent with the above mentioned
determinations at shorter wavelengths.

The present work has several aims. First of all, we used the data
set of \citet{sani11} and completed their study, by analyzing the
other known scaling relations that involve bulge properties:
kinetic energy $M_{\bullet} - M_{\star} \sigma^2$, momentum
parameter $M_{\bullet} - M_{\star} \sigma$ \citep{soker10}, and
the $M_{\bullet} - R_{\mathrm{e}} \sigma^3$ law. In order to have
a comprehensive study, we also reanalyzed the relations already
investigated by \citet{sani11}. This allows us to have a rapid
comparison among the various relationships and to find what is the
tightest.

Secondly, we examined if a simple one--to--one correlation
between, e.g., $M_{\bullet}$ and $\sigma$ is an exhaustive
description of the \citet{sani11} data, or if we have to consider
an additional dependence on a second parameter such as
$R_{\mathrm{e}}$ or $M_{\star}$. In order to study the existence
of such a dependence, we used both the approach of
\citet{marconi03}, who investigated the correlation of the
residual of the $M_{\bullet} - \sigma$ relation with
$R_{\mathrm{e}}$ and $M_{\star}$, and that of \citet{hopkins07a},
who considered the correlations between residuals at fixed
$\sigma$.

Finally, by considering the same Sloan Digital Sky Survey (SDSS)
data set, used by \citet{schawinski10} to study the role of SMBH
growth in the evolution of normal and active galaxies, we discuss
the possible consequences of the use of the scaling law
$M_{\bullet} - M_{\star} \sigma^2$ in the place of $M_{\bullet} -
\sigma$ in inferring the SMBH masses.

The paper is structured as follows. In \S \ref{Sec_2} we describe
the galaxy sample examined and the fitting procedures performed to
find the best--fitting lines for each of the relationships
considered. In \S \ref{Sec_3} we reported the fitting parameters
and showed the plots of the various scaling relations. The issue
related the existence of a SMBH fundamental plane is tackled in \S
\ref{Sec_4}. In order to emphasize how the conclusions could be
different using a scaling law rather than another one, in \S
\ref{Sec_5} we remake the analysis performed by
\citet{schawinski10} on SDSS data. After that we draw our
conclusions in \S \ref{Sec_6}.

%
%%%%%%%%%%%%%%%%%%%%%%%%%%%%%%%%%%%%%%%%%%%%%%%%%%%%%%
\section{Data analysis}
\label{Sec_2}
%%%%%%%%%%%%%%%%%%%%%%%%%%%%%%%%%%%%%%%%%%%%%%%%%%%%%%
An important fact that emerges from the \textit{Spitzer}/IRAC
images, analyzed by \citet{sani11}, is that the 3.6 $\mu$m
luminosity proved to be a very good tracer of stellar mass.
Actually, this is a crucial information in order to perform a
proper 2D photometric decomposition of the galaxy components
(disk, bar, bulge, etc.), which is useful to investigate the
interplay between central SMBHs and the evolutionary state,
luminosity and dynamics of their host galaxies.

Thanks to this bulge--disk decomposition, the authors were also
able to identify in their sample 9 disk galaxies that host a
pseudobulge\footnote{Essentially, a pseudobulge is a bulge that
shows photometric and kinematic evidence for disk--like dynamics
\citep{kormendy93}.}, i.e. Circinus, IC2560, NGC1068, NGC3079,
NGC3368, NGC3489, NGC3998, NGC4258, and NGC4594. Constructing the
correlation between $M_{\bullet}$ and the bulge 3.6 $\mu$m
luminosity $L_{3.6,\mathrm{bul}}$, \citet{sani11} noticed that
four of these nine are consistent with classical--bulge galaxies,
whereas the other 5 are outliers at more than 4$\sigma$ below
their linear regression. These galaxies are Circinus, IC2560,
NGC1068, NGC3079, and NGC3368. This result is consistent with the
fact that the $M_{\bullet}-\sigma$ relation for pseudobulges is
different from the relation in the classical bulges at a
significance level $>3\sigma$ \citep{hu08}, and that at a fixed
bulge mass, $M_{\bullet}$ in pseudobulges are on average more than
one magnitude smaller than the ones in classical bulges
\citep{hu09}. Moreover the elliptical-only galaxies, and the
non-barred galaxies, define tighter relations with less scatter
and a reduced slope than the one obtained when using a full galaxy
sample \citep{graham08b,graham2011}. For these reasons, we
neglected from the sample of \citet{sani11} the above mentioned 5
galaxies and thus considered a more consistent sample of $N=52$
galaxies, which is therefore formed by 24 ellipticals, 3 dwarf
ellipticals (dEs), 11 lenticulars, 6 barred lenticulars, 4
spirals, and 4 barred spirals. On the contrary the fits in the
paper of \citet{sani11} considered only 48 objects excluding all
the nine pseudobulges.

By using this sample (all the parameters are reported on Table 2
and 3 of \citet{sani11}), we investigated what the relationship
that best predicts the black hole mass is. In particular, the
relations that we want to study can be written in the following
form

%
% Equation 01
\begin{equation}
\log_{10}M_{\bullet} = b+m \log_{10}x, %
\label{Eq_01}
\end{equation}
where $m$ is the slope, $b$ is the normalization, and $x$ is a
parameter of the host bulge. Equation (\ref{Eq_01}) can be used to
predict the values of $M_{\bullet}$ in other galaxies once we know
the value of $x$. In order to minimize the scatter in the quantity
to be predicted, we have to perform an ordinary least-squares
regression of $M_{\bullet}$ on $x$ for the considered galaxies, of
which we already know both the quantities. We considered error
bars in both variables and, to simplify the analysis, we make all
of them symmetric about the preferred value by averaging the size
of the upper and lower 1 $\sigma$ error bars so that $x_{-l}^{+h}$
becomes $x \pm (h + l)/2$. To obtain the parameters of the fits
($m$ and $b$), we adopt the following three different fitting
methods.

\begin{itemize}
\item[$1)$]
The linear regression routine FITEXY \citep{press92} for the
relation $y=b+m x$, by minimizing the $\chi^2$
%
% Equation 02
\begin{equation}
\chi^2 = \sum_{i=1}^{N} \frac{(y_i -b-m x_i)^2}{(\Delta y_i)^2 + m^2 (\Delta x_i)^2}.%
\label{Eq_02}
\end{equation}
The most efficient and unbiased estimate of the slope is obtained
when the fitting method incorporates the residual variance, also
known as intrinsic scatter $\varepsilon_{0}$, which is that part
of the variance which cannot be attributed to specific causes
\citep{novak06}. So, if the reduced
$\chi^2_{\mathrm{r}}=\chi^2/(N-2)$ of the fit is not equal to 1,
we normalize including the suitable value of $\varepsilon_{0}$ in the Eq.(\ref{Eq_02})
to obtain
%
% Equation 03
\begin{equation}
\chi^2_{\mathrm{r}} = \frac{1}{N-2} \sum_{i=1}^{N} \frac{(y_i -b-m
x_i)^2}{(\Delta y_i)^2 + \varepsilon_{0}^2+ m^2 (\Delta x_i)^2} =
1.%
\label{Eq_03}
\end{equation}
Lastly, it is possible to obtain an estimate of the $1\sigma$
error bar on $\varepsilon_{0}$ by adjusting it until the
$\chi^2_{\mathrm{r}}$ is equal to $1+(2/N)^{1/2}$.

\item [$2)$] The linear regression FITEXY method as modified by
\citet{tremaine02}, where the measurement errors of the dependent
variable and the intrinsic scatter are added in quadrature,
adjusting $\varepsilon_{0}$ and refitting until the reduced
$\chi^2$ of the fit is equal to 1.

\item [$3)$] The Bayesian linear regression routine LINMIX\_ERR \citep{kelly07}
to determine the slope, the normalization, and the intrinsic
scatter of the relationship
\begin{equation}
\log_{10}M_{\bullet}=b+m\log_{10}(x)+\varepsilon_{0}
\label{Eq_03.1}.
\end{equation}
This routine approximates the distribution of the independent
variable as a mixture of Gaussians, bypassing the assumption of a
uniform prior distribution on the independent variable, which is
used in the derivation of $\chi^2$--FITEXY minimization routine.
Since a direct computation of the posterior distribution is too
computationally intensive, random draws from the posterior
distribution are obtained using a Markov Chain Monte Carlo method
\citep{kelly07}.
\end{itemize}

In order to be consistent with the results reported by
\citet{sani11}, we use the LINMIX\_ERR as the favorite routine to
calculate the fitting parameters of the SMBH–-bulge scaling
relations.

In Eq.(\ref{Eq_01}) in place of $x$ we considered the following
quantities: $\sigma$, $M_{\mathrm{dyn}}$, $M_{\star}$,
$M_{\mathrm{dyn}}\sigma$, $M_{\star}\sigma$,
$M_{\mathrm{dyn}}\sigma^2$, $M_{\star}\sigma^2$,
$R_{\mathrm{e}}\sigma^3$. $M_{\mathrm{dyn}}$ is the bulge
dynamical mass, which \citet{sani11} assumed dominated by stellar
matter with a negligible contribution of dark matter and gas, and
computed as:
%
% Equation 04
\begin{equation}
M_{\mathrm{dyn}} = k R_{\mathrm{e}} \sigma^2/ G,%
\label{Eq_04}
\end{equation}
where $G$ is the gravitational constant, while the factor $k$ was
fixed by the authors to be equal to 5, in agreement with
\citet{cappellari06}. On the contrary, the stellar mass
$M_{\star}$ of each galaxy has been estimated by combining the
bulge 3.6 $\mu$m luminosity with the galaxy mass--to--light ratio
($M/L$); the values are reported in Table \ref{Table_1} (Sani \&
Marconi, private communication). We follow \citet{sani11} in this
choice, even if we would have preferred an estimate of masses by
means of the Jeans equation or Schwarzschild methods as in
\citep{feoli09}, because the use of Eq.(\ref{Eq_04}) makes the the
relations $M_{\bullet}-M_{\mathrm{dyn}}\sigma$ and
$M_{\bullet}-R_{\mathrm{e}}\sigma^3$, which are deduced from
different theoretical contexts, practically equivalent.

% Table 1
\begin{table}
\caption{The values of the stellar mass and 1$\sigma$ error for each galaxy of the sample (Sani \& Marconi 2011, private communication.)}%
\label{Table_1} \centering
\begin{tabular}{c c} % columns
\hline\hline
Galaxy & $\log{M_{\star}} (+ , -)$ \\
\hline %
Circinus    &   10.26 (0.020  , 0.023)        \\
IC1459      &   12.06 (0.130  , 0.160)        \\
IC2560      &   10.94 (0.260  , 0.570)        \\
IC4296      &   12.13 (0.190  , 0.320)        \\
NGC221      &   8.814 (0.230  , 0.450)        \\
NGC524      &   11.74 (0.080  , 0.095)        \\
NGC821      &   11.50 (0.100  , 0.140)        \\
NGC1023     &   10.62 (0.130  , 0.160)        \\
NGC1068     &   11.33 (0.023  , 0.023)        \\
NGC1300     &   10.37 (0.140  , 0.190)        \\
NGC1316     &   11.96 (0.047  , 0.047)        \\
NGC2549     &   10.07 (0.059  , 0.071)        \\
NGC2748     &   10.18 (0.035  , 0.047)        \\
NGC2778     &   9.766 (0.071  , 0.071)        \\
NGC2787     &   9.980 (0.210  , 0.380)        \\
NGC2974     &   11.14 (0.059  , 0.071)        \\
NGC3031     &   11.00 (0.130  , 0.170)        \\
NGC3079     &   11.02 (0.035  , 0.035)        \\
NGC3115     &   10.99 (0.047  , 0.047)        \\
NGC3227     &   10.90 (0.190  , 0.320)        \\
NGC3245     &   10.49 (0.083  , 0.100)        \\
NGC3368     &   10.74 (0.047  , 0.047)        \\
NGC3377     &   10.63 (0.083  , 0.095)        \\
NGC3379     &   11.20 (0.071  , 0.083)        \\
NGC3384     &   10.12 (0.035  , 0.035)        \\
NGC3414     &   11.02 (0.190  , 0.320)        \\
NGC3489     &   10.07 (0.170  , 0.410)        \\
NGC3585     &   11.17 (0.230  , 0.450)        \\
NGC3607     &   11.46 (0.100  , 0.130)        \\
NGC3608     &   11.26 (0.095  , 0.110)        \\
NGC3998     &   10.36 (0.023  , 0.023)        \\
NGC4026     &   10.43 (0.083  , 0.095)        \\
NGC4151     &   10.67 (0.160  , 0.240)        \\
NGC4258     &   10.98 (0.035  , 0.035)        \\
NGC4261     &   11.63 (0.071  , 0.071)        \\
NGC4374     &   11.87 (0.023  , 0.023)        \\
NGC4459     &   10.70 (0.047  , 0.047)        \\
NGC4473     &   11.87 (0.023  , 0.023)        \\
NGC4486     &   11.94 (0.023  , 0.023)        \\
NGC4486A    &   10.06 (0.035  , 0.035)        \\
NGC4552     &   11.13 (0.035  , 0.035)        \\
NGC4564     &   10.68 (0.059  , 0.071)        \\
NGC4594     &   11.23 (0.059  , 0.071)        \\
NGC4596     &   10.81 (0.023  , 0.023)        \\
NGC4621     &   11.49 (0.047  , 0.047)        \\
NGC4649     &   11.69 (0.059  , 0.071)        \\
NGC4697     &   11.33 (0.035  , 0.047)        \\
NGC5077     &   11.75 (0.100  , 0.140)        \\
CenA        &   11.31 (0.110  , 0.150)        \\
NGC5576     &   11.39 (0.071  , 0.071)        \\
NGC5813     &   11.93 (0.083  , 0.095)        \\
NGC5845     &   10.64 (0.011  , 0.011)        \\
NGC5846     &   11.57 (0.023  , 0.023)        \\
NGC6251     &   12.36 (0.083  , 0.095)        \\
NGC7052     &   12.06 (0.071  , 0.083)        \\
NGC7457     &   9.635 (0.300  , 0.910)        \\
NGC7582     &   11.15 (0.170  , 0.270)        \\
\hline
\end{tabular}
\end{table}
%

%%%%%%%%%%%%%%%%%%%%%%%%%%%%%%%%%%%%%%%%%%%%%%%%%%%%%%
\section{Results}
\label{Sec_3}
%%%%%%%%%%%%%%%%%%%%%%%%%%%%%%%%%%%%%%%%%%%%%%%%%%%%%%
Referring to the sample of 52 galaxies, in Table \ref{Table_2} we
collected the parameters of the fits obtained thanks to the three
above mentioned fitting methods for the various relations that we
analyzed, together with the corresponding values of the $\chi^2$,
the intrinsic scatter $\varepsilon_{0}$, and the Pearson linear
correlation coefficient $r$. In Figures $1-4$, we reported the
relations in log–-log plots (we associated a particular marker to
each galaxy according to its morphological type). The best-fitting
lines are also shown for each diagram.

The relations between $M_{\bullet}$ and the corresponding kinetic
energy, momentum parameter, velocity dispersion, galaxy mass and
$R_{\mathrm{e}}\sigma^3$, fitted with a Bayesian approach to
linear regression are:

% Equations
%
{\tiny{
\begin{eqnarray}
\log_{10}M_{\bullet}=(5.30 \pm 0.26)+(0.63 \pm 0.05) %
\log_{10}\left[M_{\mathrm{dyn}}\,\sigma^2/M_{\sun}\,c^2\right]; \nonumber \\ %
(\varepsilon_{0}=0.30 \pm 0.16)\nonumber
\end{eqnarray}
\begin{eqnarray}
\log_{10}M_{\bullet}=(5.18 \pm 0.27)+(0.65 \pm 0.06) \times%
\log_{10}\left[M_{\star}\,\sigma^2/M_{\sun}\,c^2\right]; \nonumber \\ %
(\varepsilon_{0}=0.30 \pm 0.16)\nonumber%
\end{eqnarray}
\begin{eqnarray}
\log_{10}M_{\bullet}=(8.29 \pm 0.05)+(3.95 \pm 0.31)\times%
\log_{10}\left[\sigma/200 \,\mathrm{km\,s^{-1}}\right]; \nonumber \\ %
(\varepsilon_{0}=0.31 \pm 0.16),\nonumber%
\end{eqnarray}
\begin{eqnarray}
\log_{10}M_{\bullet}=(3.15 \pm 0.47)+(0.71 \pm 0.07)\times%
\log_{10}\left[R_{\mathrm{e}}\,\sigma^{3}/c\,G\,M_{\sun}\right]; \nonumber \\ %
(\varepsilon_{0}=0.33 \pm 0.17),\nonumber%
\end{eqnarray}
\begin{eqnarray}
\log_{10}M_{\bullet}=(2.64 \pm 0.53)+(0.71 \pm 0.07)\times%
\log_{10}\left[M_{\mathrm{dyn}}\,\sigma/M_{\sun}\,c\right]; \nonumber \\ %
(\varepsilon_{0}=0.33 \pm 0.17),\nonumber%
\end{eqnarray}
\begin{eqnarray}
\log_{10}M_{\bullet}=(2.47 \pm 0.54)+(0.73 \pm 0.07)\times%
\log_{10}\left[M_{\star}\,\sigma/M_{\sun}\,c\right]; \nonumber \\ %
(\varepsilon_{0}=0.34 \pm 0.17),\nonumber%
\end{eqnarray}
\begin{eqnarray}
\log_{10}M_{\bullet}=(8.18 \pm 0.06)+(0.80 \pm 0.09)\times%
\left(\log_{10}\left[M_{\mathrm{dyn}}/M_{\sun}\right]-11\right); \nonumber \\ %
(\varepsilon_{0}=0.38 \pm 0.19),\nonumber%
\end{eqnarray}
\begin{eqnarray}
\log_{10}M_{\bullet}=(8.15 \pm 0.06)+(0.80 \pm 0.09)\times%
\left(\log_{10}\left[M_{\star}/M_{\sun}\right]-11\right); \nonumber \\ %
(\varepsilon_{0}=0.40 \pm 0.19).\nonumber%
\end{eqnarray}
}}

By inspection of Table \ref{Table_2}, it is possible to note that
all the fitting parameters are consistent with previous
determinations from the literature at shorter wavelengths. The 3.6
$\mu$m $M_{\bullet}-M_{\mathrm{dyn}}\sigma^2$ relation looks to be
slightly preferable compared with the other
$M_{\bullet}-\mathrm{bulge}$ relations, especially if one
considers the values of $\chi^2_{r}$ estimated by FITEXY (column
5). However, since all correlations have similar intrinsic
scatters within the errors (column 6), we cannot conclusively
determine what the best one is.

% Table 2
\begin{table*}
\caption{Regression results for $\log M_{\bullet}=b+m \log x$ with
a sample composed by 52 galaxies. Column (1): scaling relations.
Column (2): linear regression methods. Columns (3)–-(4): the
regression coefficients, the intercept $b$ and the slope $m$.
Column (5): the reduced $\chi^2$ computed by FITEXY. Column (6):
the intrinsic scatter. Column (7):
the Pearson linear coefficient.}%
\label{Table_2} \centering
\begin{tabular}{c c c c c c c} % columns
\hline\hline Relation & Method & $b$ & $m$ & $\chi_{\mathrm{r}}^2$ & $\varepsilon_{0}$ & $r$ \\
\hline %
%---------------------------------------
$M_{\bullet}-M_{\mathrm{dyn}}\sigma^2$ & \begin{tabular}{c}
LINMIX\_ERR\\
FITEXY \\
FITEXY\_T02 \\
\end{tabular} & \begin{tabular}{c}
$5.30 \pm 0.26$ \\
$5.03 \pm 0.12$ \\
$5.31 \pm 0.24$ \\
\end{tabular} & \begin{tabular}{c}
$0.63 \pm 0.05$ \\
$0.69 \pm 0.03$ \\
$0.63 \pm 0.05$ \\
\end{tabular} & \begin{tabular}{c}
--     \\
$3.85$ \\
--     \\
\end{tabular} & \begin{tabular}{c}
$0.30 \pm 0.16$ \\
$0.30 \pm 0.04$ \\
$0.29$          \\
\end{tabular} & $0.85$ \\
\hline %
%---------------------------------------
$M_{\bullet}-M_{\mathrm{\star}}\sigma^2$ & \begin{tabular}{c}
LINMIX\_ERR\\
FITEXY \\
FITEXY\_T02 \\
\end{tabular} & \begin{tabular}{c}
$5.18 \pm 0.27$\\
$4.62 \pm 0.12$ \\
$5.22 \pm 0.24$\\
\end{tabular} & \begin{tabular}{c}
$0.65 \pm 0.06$\\
$0.77 \pm 0.03$ \\
$0.64 \pm 0.05$\\
\end{tabular} & \begin{tabular}{c}
--     \\
$5.02$ \\
--     \\
\end{tabular} & \begin{tabular}{c}
$0.30 \pm 0.16$ \\
$0.31 \pm 0.04$ \\
$0.28$ \\
\end{tabular} & $0.83$ \\
\hline %
%---------------------------------------
$M_{\bullet}-\sigma_{200}$ & \begin{tabular}{c}
LINMIX\_ERR\\
FITEXY \\
FITEXY\_T02 \\
\end{tabular} &  \begin{tabular}{c}
$8.29 \pm 0.05$\\
$8.33 \pm 0.02$ \\
$8.29 \pm 0.05$\\
\end{tabular} &  \begin{tabular}{c}
$3.95 \pm 0.31$\\
$4.77 \pm 0.13$ \\
$3.97 \pm 0.30$ \\
\end{tabular} &  \begin{tabular}{c}
--     \\
$6.03$ \\
--     \\
\end{tabular} & \begin{tabular}{c}
$0.31 \pm 0.16$ \\
$0.32 \pm 0.04$ \\
$0.29$          \\
\end{tabular} & $0.87$ \\
\hline %
%---------------------------------------
$M_{\bullet}-R_{\mathrm{e}}\sigma^3$ & \begin{tabular}{c}
LINMIX\_ERR\\
FITEXY \\
FITEXY\_T02 \\
\end{tabular} & \begin{tabular}{c}
$3.15 \pm 0.47$\\
$2.66 \pm 0.19$ \\
$3.17 \pm 0.45$\\
\end{tabular} & \begin{tabular}{c}
$0.71 \pm 0.07$\\
$0.78 \pm 0.03$ \\
$0.71 \pm 0.06$\\
\end{tabular} & \begin{tabular}{c}
--     \\
$4.92$ \\
--     \\
\end{tabular} & \begin{tabular}{c}
$0.33 \pm 0.17$ \\
$0.33 \pm 0.04$ \\
$0.32$ \\
\end{tabular} & $0.83$ \\
\hline %
%---------------------------------------
$M_{\bullet}-M_{\mathrm{dyn}}\sigma$ & \begin{tabular}{c}
LINMIX\_ERR\\
FITEXY \\
FITEXY\_T02 \\
\end{tabular} & \begin{tabular}{c}
$2.64 \pm 0.53$\\
$2.14 \pm 0.19$ \\
$2.67 \pm 0.49$ \\
\end{tabular} &  \begin{tabular}{c}
$0.71 \pm 0.07$ \\
$0.77 \pm 0.02$ \\
$0.71 \pm 0.06$ \\
\end{tabular} & \begin{tabular}{c}
-- \\
$5.10$ \\
-- \\
\end{tabular} & \begin{tabular}{c}
$0.33 \pm 0.17$ \\
$0.33 \pm 0.04$ \\
$0.32$ \\
\end{tabular} & $0.83$ \\
\hline %
%---------------------------------------
$M_{\bullet}-M_{\star}\sigma$ & \begin{tabular}{c}
LINMIX\_ERR\\
FITEXY \\
FITEXY\_T02 \\
\end{tabular} & \begin{tabular}{c}
$2.47 \pm 0.54$ \\
$1.03 \pm 0.22$ \\
$2.60 \pm 0.50$ \\
\end{tabular} & \begin{tabular}{c}
$0.73 \pm 0.07$ \\
$0.91 \pm 0.03$ \\
$0.71 \pm 0.06$ \\
\end{tabular} & \begin{tabular}{c}
--     \\
$6.85$ \\
--     \\
\end{tabular} & \begin{tabular}{c}
$0.34 \pm 0.17$ \\
$0.35 \pm 0.04$ \\
$0.31$ \\
\end{tabular} & $0.85$ \\
\hline %
%---------------------------------------
$M_{\bullet}-M_{\mathrm{dyn}}$ & \begin{tabular}{c}
LINMIX\_ERR\\
FITEXY \\
FITEXY\_T02 \\
\end{tabular} & \begin{tabular}{c}
$8.18  \pm 0.06$ \\
$8.17  \pm 0.02$ \\
$8.18  \pm 0.06$ \\
\end{tabular} & \begin{tabular}{c}
$0.80 \pm 0.09$  \\
$0.89 \pm 0.03$  \\
$0.80 \pm 0.08$  \\
\end{tabular} & \begin{tabular}{c}
--     \\
$5.43$ \\
--     \\
\end{tabular} & \begin{tabular}{c}
$0.38 \pm 0.19$   \\
$0.37 \pm 0.04$   \\
$0.36$            \\
\end{tabular} & $0.79$ \\
\hline %
%---------------------------------------
$M_{\bullet}-M_{\star}$ & \begin{tabular}{c}
LINMIX\_ERR\\
FITEXY \\
FITEXY\_T02 \\
\end{tabular} & \begin{tabular}{c}
$8.15 \pm 0.06$ \\
$8.07 \pm 0.02$ \\
$8.16 \pm 0.06$ \\
\end{tabular} & \begin{tabular}{c}
$0.80 \pm 0.09$ \\
$1.18 \pm 0.03$ \\
$0.78 \pm 0.08$ \\
\end{tabular} & \begin{tabular}{c}
--      \\
$12.71$ \\
--      \\
\end{tabular} & \begin{tabular}{c}
$0.40 \pm 0.19$ \\
$0.46 \pm 0.05$ \\
$0.37$          \\
\end{tabular} & $0.81$ \\
\hline
\end{tabular}
\end{table*}
%

%
% Figure 01
\begin{figure*}%
\centering
\includegraphics[width=16.cm]{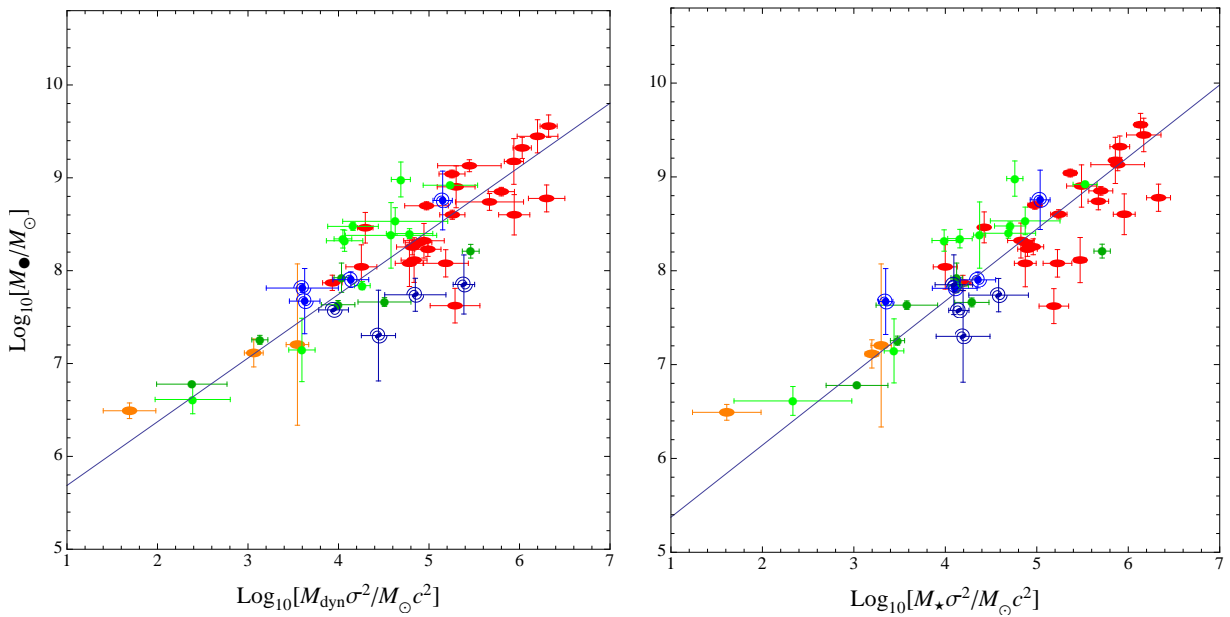}
\caption{$M_{\bullet}-M_{\mathrm{dyn}} \sigma^2$ (left) and
$M_{\bullet}-M_{\star} \sigma^2$ (right) relations for the sample
of $N=52$ galaxies extracted from the data set of \citet{sani11}.
The symbols represent elliptical galaxies (red ellipses),
lenticular galaxies (green circles), barred lenticular galaxies
(dark green circles), spiral galaxies (blue spirals), barred
spiral galaxies (dark blue barred spirals), and dwarf elliptical
galaxies (orange
ellipses). The black lines are the lines of best fit.}%
\label{Fig_01}
\end{figure*}
%
%
% Figure 02
\begin{figure*}%
\centering
\includegraphics[width=16.cm]{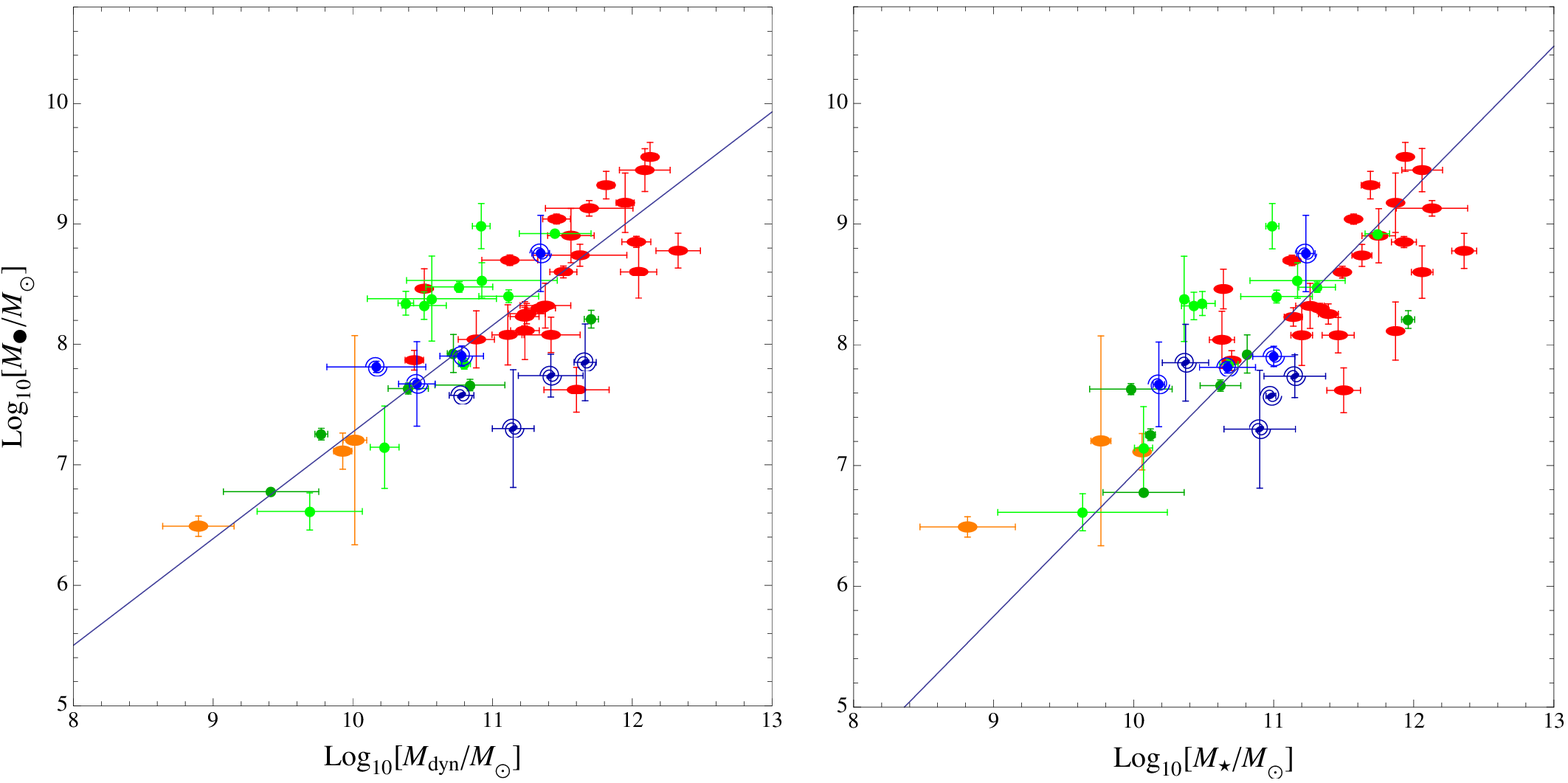}
\caption{$M_{\bullet}-M_{\mathrm{dyn}}$ (left) and
$M_{\bullet}-M_{\star}$ (right) relations for the sample of $N=52$
galaxies extracted from the data set of \citet{sani11}. The
symbols are the same as in Figure \ref{Fig_01}.}%
\label{Fig_02}
\end{figure*}
%
%
% Figure 03
\begin{figure*}%
\centering
\includegraphics[width=16.cm]{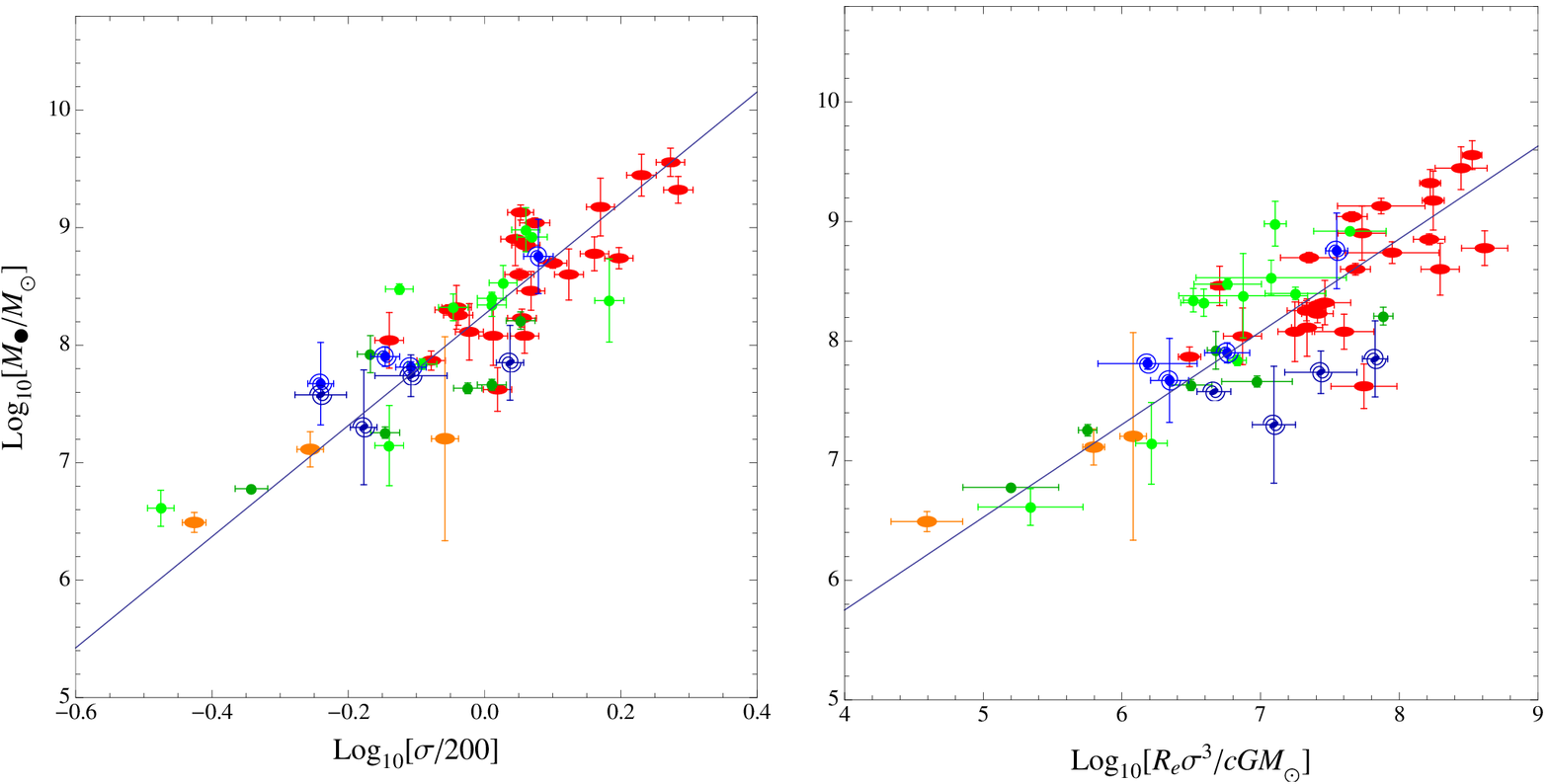}
\caption{$M_{\bullet}-\sigma_{200}$ (left) and
$M_{\bullet}-R_{\mathrm{e}}\sigma^3$ (right) relations for the
sample of $N=52$ galaxies extracted from the data set of
\citet{sani11}. The symbols are the same as in Figure \ref{Fig_01}.}%
\label{Fig_03}
\end{figure*}
%
%
% Figure 04
\begin{figure*}%
\centering
\includegraphics[width=16.cm]{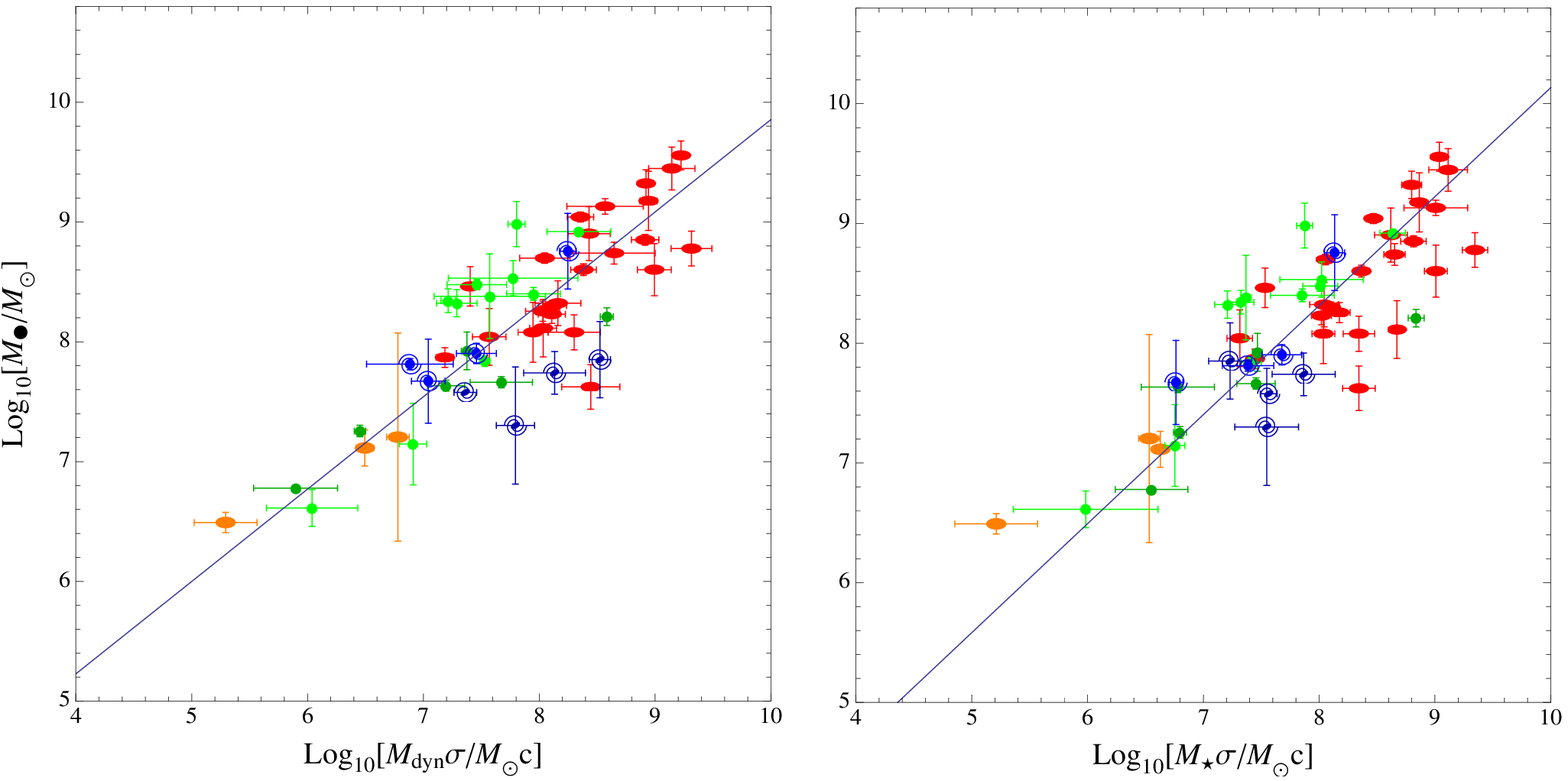}
\caption{$M_{\bullet}-M_{\mathrm{dyn}}\sigma$ (left) and
$M_{\bullet}-M_{\star}\sigma$ (right) relations for the sample of
$N=52$ galaxies extracted from the data set of \citet{sani11}. The
symbols are the same as in Figure \ref{Fig_01}.}%
\label{Fig_04}
\end{figure*}
%

%%%%%%%%%%%%%%%%%%%%%%%%%%%%%%%%%%%%%%%%%%%%%%%%%%%%%%
\section{A possible fundamental plane for supermasssive black holes}
\label{Sec_4}
%%%%%%%%%%%%%%%%%%%%%%%%%%%%%%%%%%%%%%%%%%%%%%%%%%%%%%
By analyzing a sample of 27 galaxies, which are deemed to have
``secure'' SMBH and bulge mass measurements, \citet{marconi03}
were the first to note that $M_{\bullet}$ is significantly
correlated both with $\sigma$ and with $R_{\mathrm{e}}$. Plotting
the residuals of the $M_{\bullet}-\sigma$ correlation against
$R_{\mathrm{e}}$, they concluded that a combination of $\sigma$
and $R_{\mathrm{e}}$ was necessary to drive the correlations
between $M_{\bullet}$ and other bulge properties. This topic was
then theoretically investigated by \citet{hopkins07a} by
simulations of major galaxy mergers, which defined a fundamental
plane (FP), analogous to the FP of elliptical galaxies, of the
form $M_{\bullet} \propto R_{\mathrm{e}}^{1/2} \sigma^3$ or $
M_{\bullet} \propto M^{1/2}_{\star} \sigma^2$, where $M_{\star}$
is the bulge stellar mass, and by \citet{marulli08} who found $
M_{\bullet} \propto (M_{\star} \sigma^2)^{0.7}$.\footnote{Another
effort in this sense has been performed by \citet{gultekin09b},
who analyzed the relationship among X-ray luminosity, radio
luminosity, and mass of a sample of SMBHs, identifying a FP that
can be turned into an effective SMBH mass predictor.} Moreover,
the sample of \citet{marconi03} was reanalyzed by
\citet{hopkins07b}, who found that the observations define a FP
that should be preferred over a simple relation between SMBH and
any of $\sigma$, $M_{\mathrm{dyn}}$, $M_{*}$, or $R_e$ alone at
$>3\, \sigma$ $(99.9\%)$ significance.

However, \citet{aller07} noticed that the
$M_{\bullet}-M_{\bullet}(\sigma)$ residuals for their sample of 23
galaxies did not indicate the combination suggested by
\citet{marconi03}. The evidence of a correlation between the
residuals and the effective radius is obtained by considering only
spiral and lenticular bulges. \citet{graham08b} reached the same
result studying a sample of 40 galaxies. In particular he found
that the barred galaxies are responsible for much of the trend
between the $M_{\bullet}-\sigma$ residuals and $R_{\mathrm{e}}$,
whereas the elliptical galaxies alone do not provide substantial
support for the existence of a FP plane for SMBHs. The analysis of
\citet{sani11} does not confirm the existence of a FP. In fact,
comparing the residuals of $M_{\bullet}-\sigma$ for bulges with
their effective radius, they did not found any significant
correlation either for the entire sample ($r=0.29$), or excluding
barred galaxies and/or pseudobulges ($r=0.20-0.29$).

Plotting the dependence of the residual of the
$M_{\bullet}-\sigma$ relation on $R_{\mathrm{e}}$ and $M_{*}$, the
results of \citet{hopkins07b} are clearly in conflict with that of
the other authors \citep{aller07,graham08b,sani11}. Of course, the
explanation of this difference have to be found in the different
analysis approach. In particular, \citet{hopkins07b} considered
the correlations between residuals at fixed $\sigma$, and not
simply the correlation between the residual of
$M_{\bullet}-\sigma$ and the actual value of $R_{\mathrm{e}}$ or
$M_{*}$. As stressed by \citet{hopkins07b}, if we were to do the
latter, we miss the significance of any real residuals:
``\textit{the slope recovered (i.e., the inferred dependence of
$M_{\bullet}$ on $R_{\mathrm{e}}$) is severely biased towards
being too shallow for any nonzero dependence on $R_{\mathrm{e}}$,
and in only $1\%$ of cases will such a method recover a slope
similar to the true intrinsic correlation. Looking at the
significance of the residuals in this space, it is clear that this
projection biases against detecting any significant residual
dependence on $R_{\mathrm{e}}$ or $M_{*}$}''.

\begin{figure*}%
\centering
\includegraphics[width=16.cm]{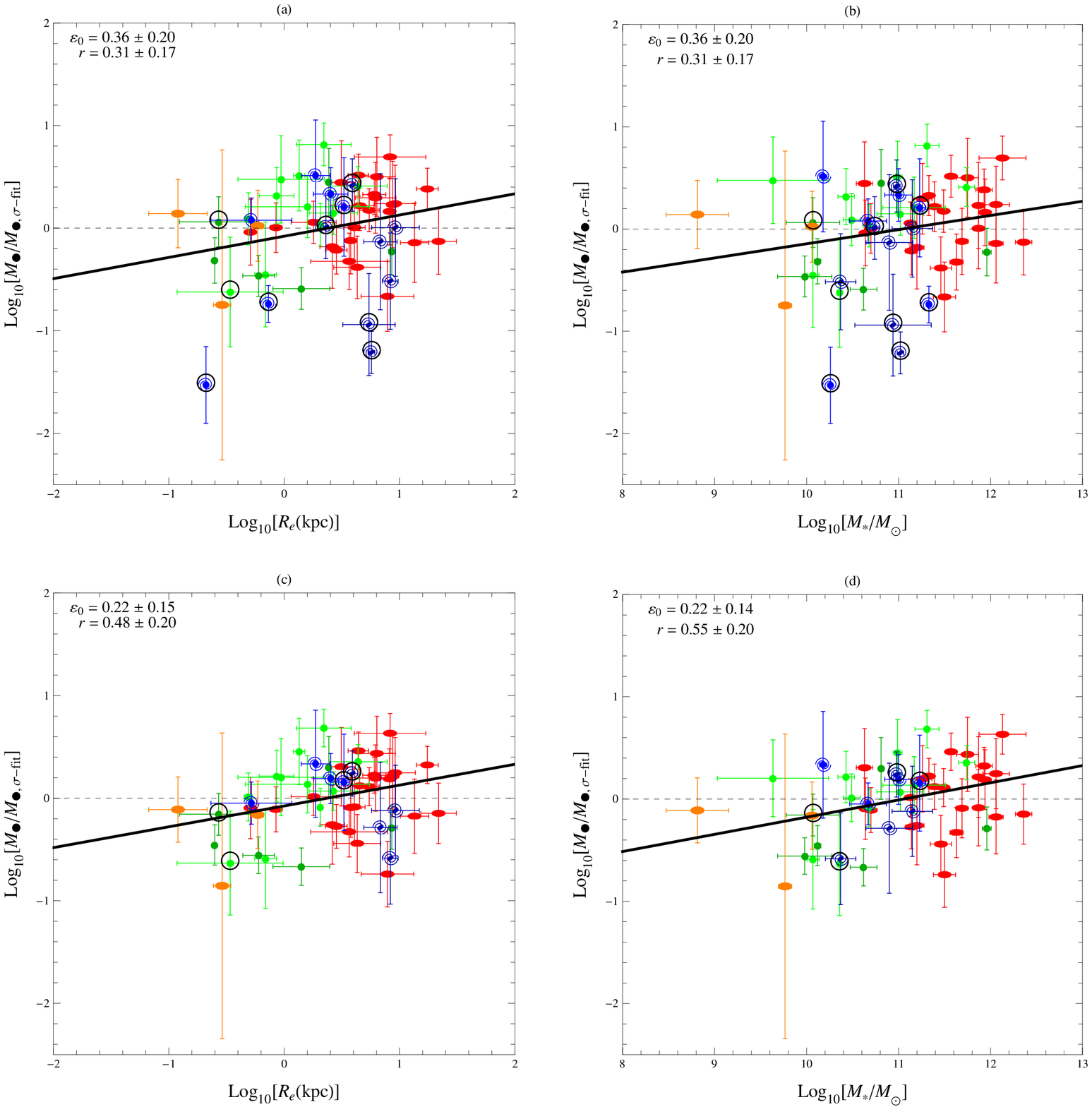}
\caption{Residuals of $M_{\bullet}-\sigma$ as a function of the
host galaxy effective radius $R_{\mathrm{e}}$ (left) and stellar
mass $M_{*}$ (right), for the full sample of 57 galaxies (a--b),
and for a more consistent sample (see text) of 52 galaxies (c--d),
respectively. The corresponding intrinsic scatter and Pearson
linear coefficient are reported in the upper--left corner of each
plot. The symbols are
the same as in Figure \ref{Fig_01}. Galaxies inside the black circles are pseudo-bulges.}%
\label{Fig_05}
\end{figure*}

\begin{figure*}%
\centering
\includegraphics[width=16.cm]{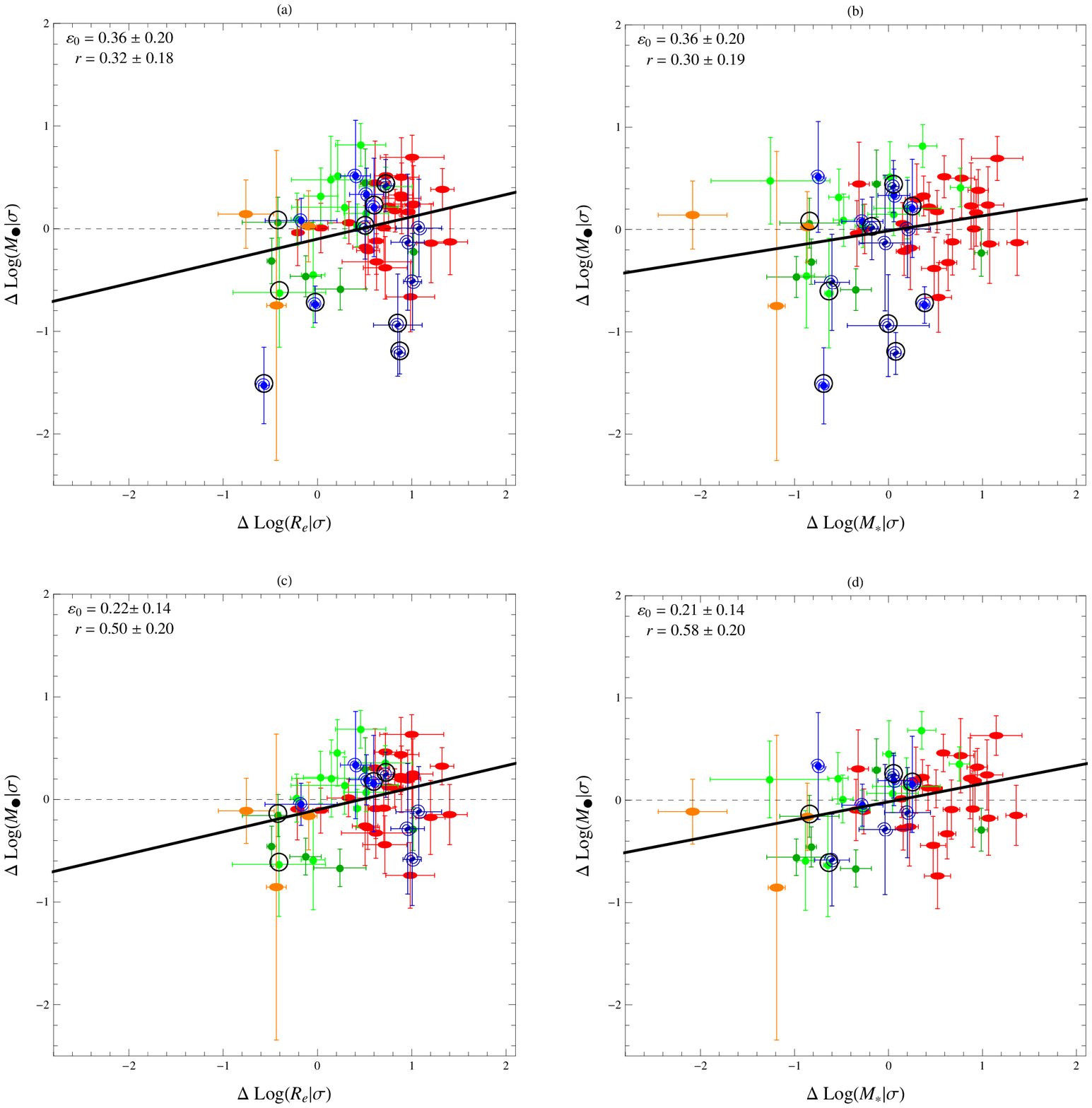}
\caption{Correlations between the residuals in the
$M_{\bullet}-\sigma$ and $R_{\mathrm{e}}-\sigma$ (left), and
$M_{*}-\sigma$ (right) relations at each $\sigma$
\citep{hopkins07b}, for the full sample of 57 galaxies (a--b), and
for a more consistent sample (see text) of 52 galaxies (b--c). The
corresponding intrinsic scatter and Pearson linear coefficient are
reported in the upper--left corner of each plot. The symbols are
the same as in Figure \ref{Fig_01}. Galaxies inside the black circles are pseudo-bulges.}%
\label{Fig_06}
\end{figure*}

Here we used both approaches performing the analysis on both the
57 and 52 galaxy samples. The results are reported in Fig.
\ref{Fig_05} and Fig. \ref{Fig_06}, where we highlighted the
position of pseudo-bulges, together with the values of the
intrinsic scatter and the Pearson linear coefficient. The
best--fitting lines have been obtained through the LINMIX\_ERR
routine.
%
% Figure 05
%
% Figure 06
In particular, Fig. \ref{Fig_05} shows the results obtained using
the \citet{marconi03} approach, whereas Fig. \ref{Fig_06} shows
the results obtained using the \citet{hopkins07b} approach; Fig.
\ref{Fig_05}a,b and \ref{Fig_06}a,b refer to the full
\citet{sani11} sample, whereas Fig. \ref{Fig_05}c,d and
\ref{Fig_06}c,d refer to the most consistent sample of 52
galaxies.

Both the approaches returned similar results, but it is
interesting to note how the values of $\varepsilon_{o}$ and $r$
slightly improve moving from the full sample to that of 52
galaxies. This fact suggests that the choice of the galaxy sample
is critical in order to get reasonable results. However, the
results reported in Table \ref{Table_2} and the above analysis of
the residuals, applied to the 52 galaxy sample, are not decisive
to confirm the result of \citet{hopkins07b}, that is the
$M_{\bullet}-M_{*}\sigma^2$ is preferred over a simple relation
between $M_{\bullet}$ and any of $\sigma$ or $M_{*}$ alone.

\section{Inferring the mass of black holes indirectly in high--redshift galaxies}
\label{Sec_5}
%%%%%%%%%%%%%%%%%%%%%%%%%%%%%%%%%%%%%%%%%%%%%%%%%%%%%%
Scaling relations between astrophysical quantities always hide
fundamental driving mechanisms. An important step, in order to
understand these physical mechanisms, is to identify where the
scaling laws apply and their nature. Without this information it
is hard to say what the best scaling law, linking the mass of the
SMBHs with the right parameter of the hosting bulges, is. For
instance, if we use a whatever scaling law in order to infer the
masses of the SMBHs located in the center of high--redshift
galaxies, and then we use them to study galaxy--evolution trends,
we could draw incorrect or misleading statements. From this point
of view, the case of the paper of \citet{schawinski10} is
emblematic. These authors used SDSS data and visual classification
of morphology from the Galaxy Zoo
project\footnote{www.galaxyzoo.org} to study black hole growth in
the nearby Universe. They selected all galaxies with SDSS spectra
classified as GALAXY \citep{strauss02} in the redshift interval
$0.02<z<0.05$; from this parent sample of $47675$ they selected a
small ($\sim 2\%$) sub sample of $942$ narrow--line Active
Galactic Nuclei (AGN), excluding broad--line AGN, highly obscured
AGN, and LINERs. Then they inferred the masses of the SMBHs
indirectly from the stellar velocity dispersion at the effective
radius via the $M_{\bullet}-\sigma$ relation using the slope and
the normalization of \citet{tremaine02}, ($b=8.13$, $m=4.02$).
Finally, they reported the distribution of inferred SMBH masses,
plotting only objects where the measured $\sigma$ was greater than
40 km s$^{-1}$, corresponding to $\log_{10}(M_{\bullet})\sim5.3$
\citep{tremaine02}.

\begin{figure*}%
\centering
\includegraphics[width=16.cm]{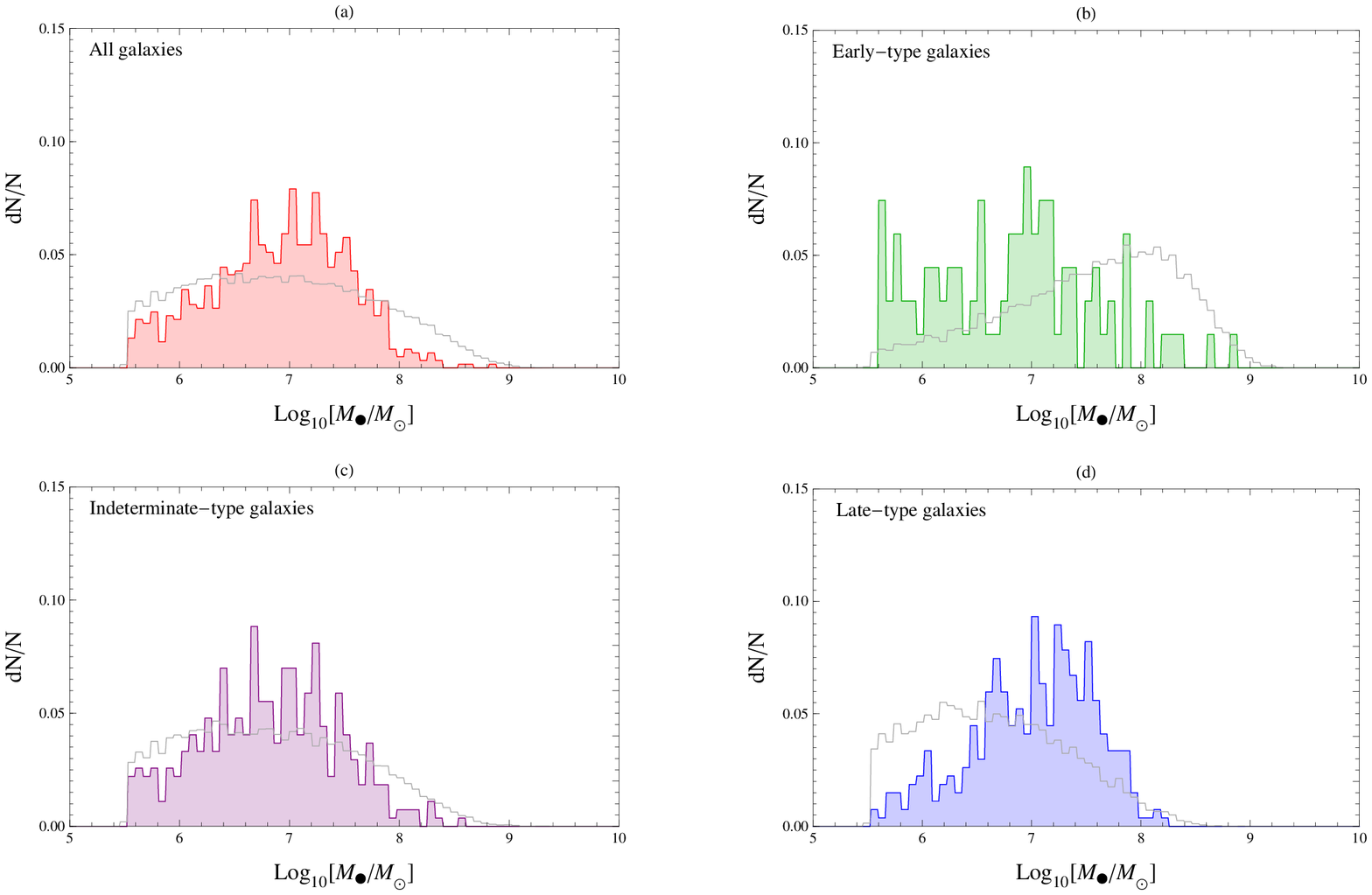}
\caption{The distribution of SMBH masses for both normal (white)
and AGN (colored) host galaxies as a function of morphology
($\Delta[\log_{10} M_{\bullet}]=0.05$ bin). The data of the
galaxies have been extracted from the Sloan Digital Sky Survey in
redshift interval $0.02<z<0.05$. The masses have been inferred via
the $M_{\bullet}-\sigma$ relation using the slope and the
normalization taken from Table \ref{Table_2}. In order to compare
our results with that of \citet{schawinski10}, we plotted
only objects where the measured velocity dispersion is greater than 40 km s$^{-1}$.} %
\label{Fig_08}
\end{figure*}

\begin{figure*}%
\centering
\includegraphics[width=16.cm]{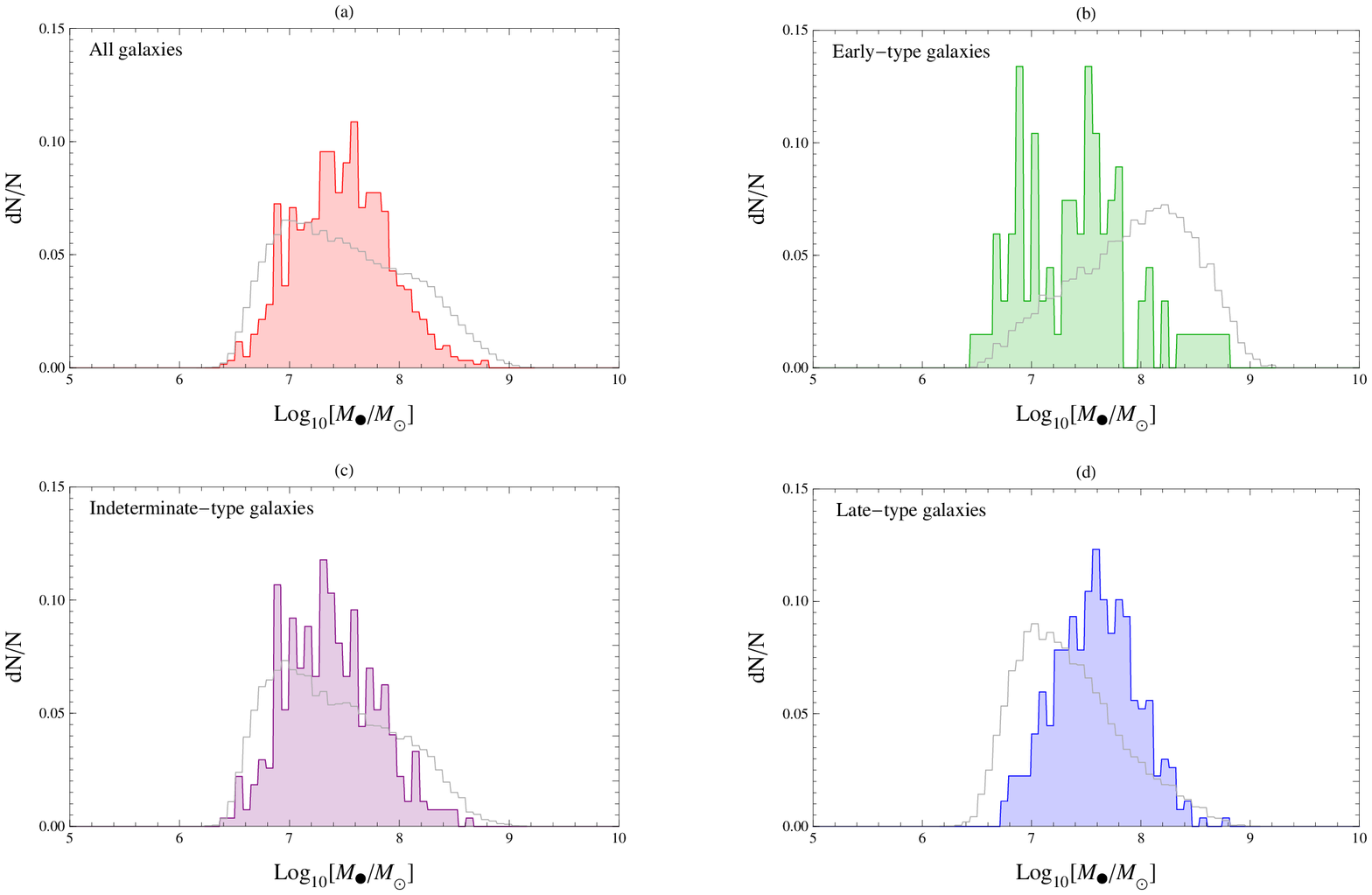}
\caption{The distribution of SMBH masses for both normal (white)
and AGN (colored) host galaxies as a function of morphology
($\Delta[\log_{10} M_{\bullet}]=0.05$ bin). The data of the
galaxies have been extracted from the Sloan Digital Sky Survey in
redshift interval $0.02<z<0.05$. The masses have been inferred
from the kinetic energy of random motions via the
$M_{\bullet}-M_{\mathrm{dyn}}\sigma^2$ relation using the slope
and the normalization taken from Table \ref{Table_2}. In order to
compare our results with that of \citet{schawinski10}, we plotted
only objects where the measured velocity dispersion is greater than 40 km s$^{-1}$.}%
\label{Fig_09}
\end{figure*}

Using the same galaxy catalogue (Table 3 of \citet{schawinski10},
and a private communication from Schawinski, 2011), we inferred
the masses of the corresponding SMBHs indirectly both from the
velocity dispersion (via the $M_{\bullet}-\sigma$ relation) and
from the kinetic energy of random motions (via the
$M_{\bullet}-M_{\mathrm{dyn}}\sigma^2$ relation), using the slopes
and normalizations obtained by the LINMIX\_ERR routine and
reported in Table \ref{Table_2}. In Fig. \ref{Fig_08} and Fig.
\ref{Fig_09} we plotted the distributions of inferred SMBH masses
from both the AGN (colored) and the normal galaxy (white)
populations, both for the entire sample and split by host
morphology. These histograms clearly highlight the different
distributions of black hole growth in agreement with the parent
population and with the adopted scaling law.

% Figure 07
%

%
% Figure 09

Comparing our Fig. \ref{Fig_08} with Fig. 8 of
\citet{schawinski10}, we find the same picture: AGN early--type
galaxies have lower black hole masses than normal early--type
galaxies, whereas for late--type galaxies there is an opposite
trend. In early-type galaxies, it is preferentially the galaxies
with the least massive black holes that are active; instead, in
late--type galaxies, it is preferentially the most massive black
holes that are active. The median values of the SMBH masses of
both early-- and late--type AGN host galaxies are still very
similar to each other ($8.38 \times 10^{6} M_{\sun}$ and $1.17
\times 10^{7} M_{\sun}$), but higher with respect to the values
($2.81 \times 10^{6} M_{\sun}$ and $4.27 \times 10^{6} M_{\sun}$)
reported by \citet{schawinski10} due to the different slope used
in the two analysis.

On the other hand, if we compare our Fig. \ref{Fig_09} with Fig.
\ref{Fig_08} or Fig. 8 of \citet{schawinski10}, it is possible to
observe the same trend of the SMBH activity in early-- and
late--type galaxies, but the distributions are much more crowded
and peaked. As a matter of fact, we found all the distribution
peaks and cut off at greater values of the SMBH mass (the median
values of the SMBH masses of early-- and late--type AGN host
galaxies are $2.85 \times 10^{7} M_{\sun}$ and $3.95 \times 10^{7}
M_{\sun}$) respectively.

Another different characteristic, which is quite evident, is that
the early--type AGN galaxies have a two--peak distribution (see
upper--right panel in Fig. \ref{Fig_09}). This feature is still
slightly present in the indeterminate--type AGN (Fig.
\ref{Fig_09}c), whereas it disappears in the late--type AGN (Fig.
\ref{Fig_09}d). It looks like there were two different kinds of
early--type AGN: one equipped with black holes of low mass (around
$3.2 \times 10^{6} M_{\sun}$), and the other one with black holes
of larger mass (around $2.5 \times 10^{7} M_{\sun}$).

Actually, a similar double-peak figure has been found by
\citet{capetti06} plotting the distributions of radio--quiet and
radio--loud AGN galaxies selected from the HST and Chandra
archival data. So the double peak in the early--type AGN (Fig.
\ref{Fig_09}b) could be easily explained by the
radio-quiet/radio-loud dichotomy.

As already noted by \citet{chiaberge05}, all radio--loud AGN are
associated with SMBH masses $\gtrsim 10^8 M_{\sun}$, whereas most
of the radio-quiet population has lower SMBH masses. A similar
result has been recently found by \citet{baldi10} and
\citet{chiaberge11}, and is in agreement with what we found in
Fig. \ref{Fig_09}b. Since radio--galaxies are almost universally
found hosted by elliptical galaxies \citep{urry95}, this explains
why we did not see any double peak in the late--type AGN galaxies
(Fig. \ref{Fig_09}d).

It would be interesting to examine a larger sample of AGN in order
to understand if this double peak is real or caused by a too small
sampling, but this is beyond the scope of this paper. What we want
to show here is that if we use the $M_{\bullet}-M_{\mathrm{dyn}}
\sigma^2$ instead of the $M_{\bullet}-\sigma$ relation, we obtain
different SMBH mass distributions (compare Fig. \ref{Fig_08} with
\ref{Fig_09}). Consequently it is better not to go on easy
conclusions regarding the activity and the evolution of both AGN
and normal host galaxies, until we have understood what the best
scaling law able to infer correctly the SMBH masses is.

%
%
%%%%%%%%%%%%%%%%%%%%%%%%%%%%%%%%%%%%%%%%%%%%%%%%%%%%%%
\section{Conclusion}
\label{Sec_6}
%%%%%%%%%%%%%%%%%%%%%%%%%%%%%%%%%%%%%%%%%%%%%%%%%%%%%%
We analyzed different scaling laws for a consistent sample formed
by $N=52$ galaxies, which have been catalogued by \citet{sani11}
on the base of \textit{Spitzer}/IRAC 3.6 $\mu$m observations of
local Universe. The sample is formed by both early--type and
late--type galaxies. Actually, from the original sample of 57
galaxies, we removed 5 disc galaxies identified by \citet{sani11}
as hosting pseudobulges and that are non consistent with the
correlations for classical bulges. For the galaxy masses, we
considered both the dynamical mass and the stellar mass. The
results of our analysis have been reported in Table \ref{Table_2},
and Figures 1-4.

The main emerging result is that the relation between the mass of
SMBHs and the kinetic energy of random motions of the host local
galaxies appears to be a robust correlation, which could provide
the right passkey to understand the nature and evolution of the
numerous observed correlations between SMBHs and host spheroid
properties. This is in agreement with our previous studies
performed on the samples of \citet{graham08a},
\citet{gultekin09a}, and \citet{hu09}. Even if the values of the
$\chi_{\mathrm{r}}^2$  (see Table \ref{Table_2}) indicate that the
$M_{\bullet}-M_{\mathrm{G}}\sigma^2$ works better than the others,
by considering the intrinsic scatter of the various relations and
in particular their errors, we cannot conclusively determine the
best relation, because all the examined laws appear on the same
level. The comprehensive analysis of residuals discussed in \S
\ref{Sec_4} does not confirm the result claimed by
\citet{hopkins07b}, according to which a black hole FP should be
preferred over a simple one--one relation, even if it cannot be
definitively ruled out.

%This is in agreement with our previous studies performed on the
%samples of \citet{graham08a}, \citet{gultekin09a}, and
%\citet{hu09}. However, by considering the intrinsic scatter of the
%various relations and in particular their errors, we do not have
%any conslusive indication that the
%$M_{\bullet}-M_{\mathrm{G}}\sigma^2$ works better than the others.
%The comprehensive analysis of residuals discussed in \S
%\ref{Sec_4} does not confirm the result claimed by
%\citet{hopkins07b} that a black hole FP should be preferred over a
%simple one--one relation, even if it cannot be definitively ruled
%out.

Since it has now been tested on four different samples of
galaxies, independently catalogued by different authors, the
goodness of the $M_{\bullet}-M_{\mathrm{G}} \sigma^2$ relation as
a predictor of the SMBH mass in the center of galaxies is enough
robust. Again, this relation is the only one that currently has a
quite clear physical explanation. In this perspective, as we
discussed in \S \ref{Sec_5}, in order to obtain correct estimates
of SMBH masses, the $M_{\bullet}-M_{\mathrm{G}} \sigma^2$ relation
should be preferably used instead of the other popular scaling
laws. As a matter of fact, the SMBH mass distribution of the
early--type AGN galaxies, inferred by a much more physically
motivated relation, clearly shows the radio--quiest/radio--loud
dichotomy. The same result is not achieved if we use the usual
$M_{\bullet}-\sigma$ relation.

\begin{acknowledgements}
We wish to thank Roberto De Carli, Alister Graham, Alessandro
Marconi, Nikolay Nikolov, Eleonora Sani, Kevin Schawinski for
their useful suggestions and private communications. We also thank
the referee for many suggestions that have helped us to improve
our paper considerably. L.M. thanks the Harvard Smithsonian Center
for Astrophysics for the kind hospitality, and acknowledges
support for this work by research funds of the University of
Sannio, and the International Institute for Advanced Scientific
Studies.
\end{acknowledgements}

\bibliographystyle{aa} % style aa.bst

\begin{thebibliography}{}
%
\bibitem[Aller \& Richstone(2007)]{aller07}
Aller, M. C., \& Richstone, D. O. 2007, \apj, 665, 120
%
\bibitem[Baes et al.(2003)]{baes03}
Baes, M., Buyle, P., Hau, G. K. T., Dejonghe 2003, \mnras, 341,
L44
%
\bibitem[Baldi \& Capetti(2010)]{baldi10}
Baldi R. D., \& Capetti A. 2010, \aap, 519, A48
%
\bibitem[Bellovary et al.(2011)]{bellovary11}
Bellovary, J., Volonteri, M., Governato, F., et al. 2011,
submitted to \apj, arXiv:1104.3858
%
\bibitem[Bower et al.(2006)]{bower06}
Bower, R. G., Benson, A. J., Malbon, R., et al. 2006, \mnras, 370,
645
%
\bibitem[Burkert \& Tremaine(2010)]{burkert10}
Burkert, A., \& Tremaine, S. 2010, \apj, 720, 516
%
\bibitem[Capetti \& Balmaverde(2006)]{capetti06}
Capetti, A., \& Balmaverde, B. 2006, \aap, 453, 27
%
\bibitem[Cappellari et al.(2006)]{cappellari06}
Cappellari, M., Bacon, R., Bureau, M., et al. 2006, \mnras, 366,
1126
%
\bibitem[Chiaberge et al.(2005)]{chiaberge05}
Chiaberge, M., Capetti, A., Macchetto, F. D. 2005, \apj, 625, 716
%
\bibitem[Chiaberge \& Marconi(2011)]{chiaberge11}
Chiaberge, M., \& Marconi, A. 2011, \mnras, 416, 917
%
\bibitem[De Lucia \& Blaizot(2007)]{delucia07}
De Lucia, G., Blaizot, J. 2007, \mnras, 375, 2
%
\bibitem[Feoli \&  Mele(2005)]{feoli05}
Feoli, A., \& Mele, D. 2005, Int. Jour. Mod. Phys. D, 14, 1861
%
\bibitem[Feoli \&  Mele(2007)]{feoli07}
Feoli, A., \& Mele, D. 2007, Int. Jour. Mod. Phys. D, 16, 1261
%
\bibitem[Feoli \& Mancini(2009)]{feoli09}
Feoli, A., \&  Mancini, L. 2009, \apj, 703, 1502
%
\bibitem[Feoli et al.(2011)]{feoli11a}
Feoli, A., Mancini, L., Marulli, F., van den Bergh, S. 2011, Gen.
Rel. Grav., 43, 1007
%
\bibitem[Feoli \& Mancini(2011)]{feoli11b}
Feoli, A., \&  Mancini, L. 2011, to appear on Int. Jour. Mod.
Phys. D, arXiv:1012.3160
%
\bibitem[Ferrarese \& Merritt(2000)]{ferrarese00}
Ferrarese, L., \&  Merritt, D. 2000, \apj, 539, L9
%
\bibitem[Ferrarese(2002)]{ferrarese02}
Ferrarese, L. 2002, \apj, 578, 90
%
\bibitem[Gebhardt et al.(2000)]{gebhardt00}
Gebhardt, K., Bender, R., Bower, G., et al. 2000, \apjl, 539, 13
%
\bibitem[Gebhardt et al.(2003)]{gebhardt03}
Gebhardt, K., Richstone, D. O., Tremaine, S., et al. 2003, \apj,
583, 92
%
\bibitem[Graham \& Driver(2005)]{graham05}
Graham, A. W., \& Driver, S. P. 2005, \pasa, 22(2), 118
%
\bibitem[Graham \& Driver(2007)]{graham07}
Graham, A. W., \& Driver, S. P. 2005, \apj, 655, 77
%
\bibitem[Graham(2008a)]{graham08a}
Graham, A. W. 2008, \pasa, 25, 167
%
\bibitem[Graham(2008b)]{graham08b}
Graham, A. W. 2008b, \apj, 680, 143
%
\bibitem[Graham(2011)]{graham11}
Graham, A. W. 2011, arXiv:1103.0525
%
\bibitem[Graham et al.(2011)]{graham2011}
Graham, A. W., Onken, C. A., Athanassoula, E.; Combes, F. 2011, \mnras, 412, 2211
%
\bibitem[G\"{u}ltekin et al.(2009a)]{gultekin09a}
G\"{u}ltekin, K., Cackett, E.~M., Miller, J.~M., et al. 2009a,
\apj, 698, 198
%
\bibitem[G\"{u}ltekin et al.(2009b)]{gultekin09b}
G\"{u}ltekin, K., Cackett, E.~M., Miller, J.~M., et al. 2009b,
\apj, 706, 404
%
\bibitem[H\"{a}ring \& Rix(2004)]{haring04}
H\"{a}ring, N., \& Rix, H. 2004, \apjl, 604, L89
%
\bibitem[Ho(2004)]{ho04}
Ho, L. C. (ed.) 2004, Coevolution of Black Holes and Galaxies
(Carnegie Observatories Astrophys. Ser. 1, Cambridge Univ. Press)
%
\bibitem[Hopkins et al.(2007a)]{hopkins07a}
Hopkins P. F., Hernquist L., Cox T. J. et al., 2007a, \apj, 669,
45
%
\bibitem[Hopkins et al.(2007b)]{hopkins07b}
Hopkins P. F., Hernquist L., Cox T. J. et al., 2007b, \apj, 669,
67
%
\bibitem[Hopkins(2008)]{hopkins08}
Hopkins, P. F. 2008, in IAU Symp. 245, Formation and Evolution of
Galaxy Bulges, ed. M. Bureau, E. Athanassoula, \& B. Barbuy
(Cambridge: Cambridge Univ. Press), 219
%
\bibitem[Hu(2008)]{hu08}
Hu, J. 2008, \mnras, 386, 2242
%
\bibitem[Hu(2009)]{hu09}
Hu, J. 2009, arXiv:0908.2028
%
\bibitem[Jahnke \& Macci\`{o}(2011)]{jahnke11}
Jahnke, K., \& Macci\`{o}, A. 2011, \apj, 734, 92
%
\bibitem[Kelly(2007)]{kelly07}
Kelly, B. C. 2007, \apj, 665, 1489
%
\bibitem[Kisaka et al.(2008)]{kisaka08}
Kisaka, S., Kojima, Y., Otani, Y. 2008, \mnras, 390, 814
%
\bibitem[Kormendy(1993)]{kormendy93}
Kormendy, J. 1993, in IAU Symposium 153: Galactic Bulges, ed. H.
DeJonghe \& H. Habbing (Kluwer: Dordrecht), p 209
%
\bibitem[Kormendy \& Richstone(1995)]{kormendy95}
Kormendy, J., \& Richstone, D. 1995, \araa, 33, 581
%
\bibitem[Kormendy \& Bender(2011)]{kormendy11}
Kormendy, J., \& Bender, R. 2011, \nat, 469, 377
%
\bibitem[Laor(2001)]{laor01}
Laor, A., 2001, \apj, 553, 677
%
\bibitem[Lauer et al.(2007)]{lauer07}%
Lauer, T. R., Faber, S. M., Richstone, D., et al. 2007, \apj, 662,
808
%
\bibitem[Marulli et al.(2008)]{marulli08}
Marulli, F., Bonoli, S., Branchini, E., et al. 2008, \mnras, 385,
1846
%
\bibitem[Novak et al.(2006)]{novak06}
Novak, G. S., Faber, S. M., Dekel, A., 2006, \apj 637, 96
%
\bibitem[Magorrian et al.(1998)]{magorrian98}
Magorrian, J., Tremaine, S., Richstone, D., et al. 1998, \aj, 115,
2285
%
\bibitem[Marconi \& Hunt(2003)]{marconi03}
Marconi, A., \& Hunt, L.~K. 2003, \apj, 589, L21
%
\bibitem[Merritt \& Ferrarese(2001)]{merritt01}
Merritt, D., \& Ferrarese, L. 2001, \apj, 547, 140
%
\bibitem[Press et al.(1992)]{press92}
Press, W. H., Teukolsky, S. A., Vetterling, W. T., Flannery, B. P.
1992, Numerical Recipes, 2nd edn. Cambridge University Press,
Cambridge
%
\bibitem[Richstone et al.(1995)]{richstone95}
Richstone, D., Ajhar, E. A., Bender, R., et al. 1998, \nat, 395,
A14
%
\bibitem[Richstone et al.(1998)]{richstone98}
Richstone, D., Ajhar, E. A., Bender, R., et al. 1998, \nat, 395,
A14
%
\bibitem[Sani et al.(2011)]{sani11}
Sani, E., Marconi, A., Hunt, L. K., Risaliti, G. 2011, \mnras,
413, 1479
%
\bibitem[Schawinski et al.(2010)]{schawinski10}
Schawinski, K., Urry, C. M., Virani, S., et al. 2010, \apj, 711,
284
%
\bibitem[Schneider(2006)]{schneider06}
Schneider, P. 2006, Extragalactic astronomy and cosmology
(Springer--Verlag Berlin Heidelberg)
%
\bibitem[Snyder et al.(2011)]{snyder11}
Snyder, G. F., Hopkins, P. F., Hernquist, L. 2011, \apj, 728, L24
%
\bibitem[Soker \& Meiron(2011)]{soker10}
Soker, N., \& Meiron, Y. 2011, \mnras, 411, 1803
%
\bibitem[Strauss et al.(2001)]{strauss02}
Strauss, M. A., Michael, A., Weinberg, D. H., et al. 2002, \aj,
124, 1810
%
\bibitem[Tremaine et al.(2002)]{tremaine02}
Tremaine, S., Gebhardt, K., Bender, R., et al. 2002, \apj, 574,
740
%
\bibitem[Urry \& Padovani(1995)]{urry95}
Urry, C. M., \& Padovani, P. 2011, \pasp, 107, 803
%
\bibitem[Volonteri et al.(2011)]{volonteri11}
Volonteri, M., Natarajan, P., Gultekin, K. 2011, \apj, 737, 50
%
\bibitem[Wandel(2002)]{wandel02}
Wandel, A. 2002, \apj, 565, 762
%
%
\end{thebibliography}

\end{document}